\documentstyle[12pt]{article}
\newcommand{\drawsquare}[2]{\hbox{%
\rule{#2pt}{#1pt}\hskip-#2pt%  left vertical
\rule{#1pt}{#2pt}\hskip-#1pt%  lower horizontal
\rule[#1pt]{#1pt}{#2pt}}\rule[#1pt]{#2pt}{#2pt}\hskip-#2pt%  upper horizontal
\rule{#2pt}{#1pt}}% right vertical
\newcommand{\Yfund}{\raisebox{-.5pt}{\drawsquare{6.5}{0.4}}}%  fund
\newcommand{\Ysymm}{\raisebox{-.5pt}{\drawsquare{6.5}{0.4}}\hskip-0.4pt%
        \raisebox{-.5pt}{\drawsquare{6.5}{0.4}}}%  symmetric second rank
\newcommand{\Ythrees}{\raisebox{-.5pt}{\drawsquare{6.5}{0.4}}\hskip-0.4pt%
          \raisebox{-.5pt}{\drawsquare{6.5}{0.4}}\hskip-0.4pt% 
          \raisebox{-.5pt}{\drawsquare{6.5}{0.4}}}%  symmetric third rank
\newcommand{\Yfours}{\raisebox{-.5pt}{\drawsquare{6.5}{0.4}}\hskip-0.4pt%
          \raisebox{-.5pt}{\drawsquare{6.5}{0.4}}\hskip-0.4pt% 
          \raisebox{-.5pt}{\drawsquare{6.5}{0.4}}\hskip-0.4pt% 
          \raisebox{-.5pt}{\drawsquare{6.5}{0.4}}}%  symmetric fourth rank
\newcommand{\Yasymm}{\raisebox{-3.5pt}{\drawsquare{6.5}{0.4}}\hskip-6.9pt%
        \raisebox{3pt}{\drawsquare{6.5}{0.4}}}%  antisymmetric second rank
\newcommand{\Ythreea}{\raisebox{-3.5pt}{\drawsquare{6.5}{0.4}}\hskip-6.9pt%
        \raisebox{3pt}{\drawsquare{6.5}{0.4}}\hskip-6.9pt
        \raisebox{9.5pt}{\drawsquare{6.5}{0.4}}}

\newcommand{\Ysquare}{\raisebox{-3.5pt}{\drawsquare{6.5}{0.4}}\hskip-0.4pt%
        \raisebox{-3.5pt}{\drawsquare{6.5}{0.4}}\hskip-13.4pt%
        \raisebox{3pt}{\drawsquare{6.5}{0.4}}\hskip-0.4pt%
        \raisebox{3pt}{\drawsquare{6.5}{0.4}}}%  4 boxes in a square
\newcommand{\Ysymtwoathree}{\raisebox{-3.5pt}{\drawsquare{6.5}{0.4}}
       \hskip-0.4pt%
        \raisebox{-3.5pt}{\drawsquare{6.5}{0.4}}\hskip-13.4pt%
        \raisebox{3pt}{\drawsquare{6.5}{0.4}}\hskip-0.4pt%
        \raisebox{3pt}{\drawsquare{6.5}{0.4}}\hskip-13.4pt
        \raisebox{9.5pt}{\drawsquare{6.5}{0.4}}\hskip-0.4pt%
        \raisebox{9.5pt}{\drawsquare{6.5}{0.4}}}% 6 boxes
\newcommand{\Yfiveboxes}{\raisebox{-3.5pt}{\drawsquare{6.5}{0.4}}\hskip-6.9pt%
        \raisebox{3pt}{\drawsquare{6.5}{0.4}}\hskip-0.4pt%
        \raisebox{3pt}{\drawsquare{6.5}{0.4}}\hskip-13.4pt
        \raisebox{9.5pt}{\drawsquare{6.5}{0.4}}\hskip-0.4pt
        \raisebox{9.5pt}{\drawsquare{6.5}{0.4}}} % 5 boxes 
\newcommand{\Ysymthreeatwo}{\raisebox{-3.5pt}{\drawsquare{6.5}{0.4}}
          \hskip-0.4pt%
          \raisebox{-3.5pt}{\drawsquare{6.5}{0.4}}\hskip-0.4pt% 
          \raisebox{-3.5pt}{\drawsquare{6.5}{0.4}}\hskip-19.9pt
          \raisebox{3pt}{\drawsquare{6.5}{0.4}}\hskip-0.4pt%
          \raisebox{3pt}{\drawsquare{6.5}{0.4}}\hskip-0.4pt% 
          \raisebox{3pt}{\drawsquare{6.5}{0.4}}}% six boxes

\def\equ#1{\begin{equation}#1\end{equation}}
\begin{document}
%\tightenlines
%
%\draft
%\twocolumn[       %typeset the title and abstract in one column
%{\tighten
%\preprint{\vbox{
%\hbox{UCSD/PTH 98--??}
%}}

\begin{titlepage}
\title{Systematic Study of Theories with Quantum Modified Moduli. II
}
\author{Benjam\'\i{}n Grinstein\cite{bg}
and Detlef R. Nolte\cite{dn}\\
Department of Physics, University of California at San Diego\\
La Jolla, CA 92093}
%\bigskip
\date{March 1998
}

\maketitle
\begin{abstract}
\vskip-4truein\hbox{\hskip4truein UCSD/PTH 98--11}\vskip4truein 
We complete the process of classifying all supersymmetric theories
with quantum modified moduli. We present all the supersymmetric gauge
theories based on a simple orthogonal or exceptional group that
exhibit a quantum modified moduli space.  The quantum modified
constraints of theories derived from s-confining theories are
invariant under all symmetries.  However, theories that cannot be
obtained by a deformation of an s-confining theory may have
constraints that are covariant, rather than invariant.
\end{abstract}

\end{titlepage}
%\pacs{11.30.Pb, 11.15.-q, 11.15.Tk}
%}% end the tighten
%
% ] % end twocolumn format
%
%\narrowtext
\section{Introduction}

There has been much recent progress in our understanding of the the
phase structure of supersymmetric gauge field theories \cite{reviews}.
Seiberg\cite{seiberg1} considered SUSY QCD with $N_c$ colors and
$N_{f}$ quarks in the fundamental and antifundamental representation
of $SU(N_c)$.  For $N_{f} \leq N_{c}+1$ Seiberg found a description of
the original theory in terms of the confined degrees of freedom which
are gauge invariant operators in the matter fields.  Using the
symmetries and the holomorphy of the superpotential he determined the
superpotential of these theories. These superpotentials can only be
generated non-perturbatively because of the non-renormalisation
theorem for the superpotential in perturbation theory.  In this range
of $N_{f}$ one can distinguish three types of theories based on the
changes made on the classical vacuum.  The superpotential for $N_{f} <
N_{c}$ lifts the classical vacuum.  For $N_{f} = N_{c}$ it just
modifies the classical vacuum. This modification appears as a quantum
change to a classical constraint between the gauge invariant degrees
of freedom. For $N_{f} = N_{c} + 1$ classical and quantum vacua are
the same. There are still constraints between the gauge invariant
operators, but they are not modified. These qualitatively different
phenomena can be looked for in theories with different matter content
and different gauge groups.  Generally, one calls ``quantum modified
theories''\cite{seiberg1,GN} those in which a classical
constraint between the gauge invariant degrees of freedom is changed
due to quantum effects.  Following Csaki et al\cite{CSS} we call the
analogs to SUSY QCD $N_{f} = N_{c} + 1$ ``s-confining'' theories.
Each type of theory needs a specific superpotential to implement such
a vacuum modification. The special features of the superpotentials are
used to classify the s-confining theories \cite{CSS} and the quantum
modified theories \cite{GN}.

In this paper we complete our work on the classification of quantum
modified theories\cite{GN}.  We do this in two different ways. First
we classify quantum modified theories for the simple gauge groups not
considered in Ref.~\cite{GN}. That means we concentrate on the SO and
exceptional ($G_2$, $F_4$, $E_{6,7,8}$) groups. We give a detailed
description of the quantum modified theories which are not derived
from s-confining theories. Second we elaborate on a claim in
\cite{GN}. We argued there that in theories with a covariantly
modified moduli space (c-QMM theories) new branches can appear on
certain points in moduli space. At the branch the Lagrange multiplier
is massless and it is therefore responsible for the branch. The
identity of the Lagrange multiplier was not known and therefore the
new branch seemed to be mysterious. We study a theory based on a
non-simple gauge group
in which the c-QMM theory is derived from a s-confining theory. It is
now possible to identify the Lagrange multiplier with a composite
field of the parent s-confining theory. By moving through the moduli
space the mass of this composite field changes. At certain points on
moduli space the mass of this particle vanishes and there the
supersymmetric vacuum has a new branch. For the  c-QMM theories based
on a simple gauge group
such an interpretation of the Lagrange multiplier is difficult to
proof because they cannot be derived from s-confining theories.  

We have found that if a simple group gauge theory derived from an
s-confining theory has a quantum modified moduli, then this moduli is
described by a quantum modified constraint that is invariant under all
symmetries. For these ``i-QMM theories'' the Lagrange multipliers are
again composite fields of the parent s-confining theories.  If there
is just one constraint then there cannot be a new branch on which the
Lagrange multiplier becomes massless. SUSY--QCD is an example of such
a theory. But if the theory has multiple constraints new branches can
occur.  We will give an example in which this happens.

In section 2 we collect the tools needed for a systematic
classification of all quantum modified theories (for details and
derivations see \cite{GN}). In this section we also list all the
flows.  A discussion of new branches found in c-QMM theories or
theories with multiple constraints can be found in section~3. Our
results for the SO and exceptional groups are presented in section~4.
Section~5 has conclusions. In an appendix we present a method which
helps in determining how theories flow. We used this method to find
the branches for the exceptional groups.

%%%%%%%%%%%%%%%%%%%%%%%%%%%%%%%%%%%%%%%%%%%%%%%%%%%%%%%%%%%%%%%%%%%%%%%%%%%%%%
\section{Index Condition and Flows}
\subsection{The Index Condition}

Quantum modified theories have at least one constraint between the 
gauge invariant operators \(\phi_{i} \). At the classical level, the
constraints are schematically of the form
\[  \sum_{n=1}^{m} (\prod_{i=1}^{k_{n}} \phi_{i})_{n} = 0 .\]
Quantum dynamics modifies this constraint to
\equ{\label{Qconst}
\sum_{n=1}^{m} (\prod_{i=1}^{k_{n}} \phi_{i} )_{n} 
= \prod_{i} \phi_{i} \Lambda^{p} .}
The important feature is the scale $\Lambda$ appearing on the right
side of Eq.~(\ref{Qconst}). If the original classical constraint is
not invariant under all symmetries, one has to multiply the
$\Lambda^{p}$ term by some gauge invariant operators which produces a
constraint covariant under all the symmetries. Generally the symmetries
do not fix the gauge invariant operator uniquely. By flowing the
covariant theories to known theories one can determine the gauge
invariant operator uniquely. We refer to these covariant theories as
c-QMM theories and for theories with invariant classical constraints as
i-QMM theories.  This distinction is physically important. We describe
this in Ref.~\cite{GN} and in section \ref{branch}.

To understand the index constraint we assign a $U(1)_{R}$ symmetry to
each matter field separately and require that both sides have the same
charge under all the $U(1)_{R}$ symmetries. Because the scale
$\Lambda$ appears one one side of the constraint there is at least one
field which appears as a higher power on the left than on the right.
Therefore the $U(1)_{R}$ charge of this field has to be
zero. Otherwise the constraint would not be covariant under this
$U(1)_{R}$ symmetry.  But gauge anomaly cancellation for this
$U(1)_{R}$ symmetry requires: 
\equ{\label{index-const}\sum_{i=1}^{n} \mu_{i} - \mu_{G} = 0 .}  
We call this the index condition. For another argument and more
details see Refs.~\cite{CSS,GN}.

%----------------------------------------------------------------------------- 
\subsection{The Flow}
%-----------------------------------------------------------------------------
\subsubsection{The Flow of the Theories}
%------------------------------------------------------------------------------

The index condition is only a necessary condition for a quantum
deformation of the moduli. To find out if a candidate theory actually
has a QMM, one must make further investigations.  As the next step to
sort out all QMM theories we consider points in the classical moduli
space where the gauge group of the candidate theory is broken. The
gauge fields which correspond to the broken generators acquire a mass
proportional to the vacuum expectation value of the Higgs field. These
massive gauge superfields pair up with chiral superfields which become
massive through the super-Higgs mechanism as well. Together they form
a massive supermultiplet. We integrate out these heavy degrees of
freedom. The new theory, which is an effective theory of the original
`UV' theory, should be in a phase consistent with the UV theory being
in a quantum modified phase.  We refer to this as `the flow' of the UV
theory to an effective theory. If the theory flows to a theory in a
Coulomb phase\cite{CS2} we say that the theory has a Coulomb branch,
not a QMM. By studying the flow we can, therefore, rule out quite a
few theories which fulfill the index condition.

It is useful to tabulate the manner in which theories may flow.  Below
we list the gauge groups together with their particle content. 
For a $SO(2N)$ theory with
$s$ spinors, $s'$ conjugate spinors and $q$ vector representations as the
matter contents we write $(s, s', q)$. For a $SO(2N+1)$ theory with
$s$ spinors and $q$ vector representations as the matter contents we
write $(s, q)$. For other theories the
matter is presented in square brackets and is represented by the Young
tableaux of the corresponding representation, with a possible
multiplier when there are more than one field for that
representation. We do not list any gauge singlets that may remain in
the effective theory. These are not all the possible flow
diagrams. They were, however, sufficient for our classification work.

\begin{eqnarray}
\label{flow1}
\mbox{Simple Gauge Group} \; [\mbox{adjoint}] & \longrightarrow & 
\mbox{Coulomb branch}
\end{eqnarray}
\begin{eqnarray}
\label{flow2}
SO(N) [(N-2)\; \Yfund] & \longrightarrow & \mbox{Coulomb branch}
\end{eqnarray}
\clearpage
\begin{eqnarray}
\label{flow3}
&SO(14) (1,0,4) &\longrightarrow G_2 \times G_2 [4(\Yfund,1)+4(1,\Yfund)]\\
&\downarrow  & \hspace{0.5in}\downarrow\nonumber\\
&SO(13) (1,3) & \longrightarrow G_2 \times SU(3)[4(\Yfund,1)+
3(1,\Yfund)+3(1,\overline{\Yfund})]\nonumber\\
&\downarrow &   \nonumber               \\
SU(6)[\Ythreea + 3(\Yfund + \overline{\Yfund})] \longleftarrow 
& SO(12) (1,1,2)  & SO(12) (2,0,2) \longrightarrow SU(6) 
[\Yasymm +\overline{\Yasymm}+ 2(\Yfund + \overline{\Yfund})]\nonumber \\
&\downarrow& \swarrow \nonumber\\
&SO(11) (2,1) & \nonumber \\
&\downarrow &\nonumber\\
&SO(10) (2,2,0) & \longrightarrow SU(5) [5(\Yfund + \overline{\Yfund})]
 \nonumber
\end{eqnarray}
\begin{eqnarray}
\label{flow4}
E_7 \;[3 \; \Yfund ] & &\nonumber\\
\downarrow   & &\nonumber\\
E_6 \;[2(\Yfund + \overline{\Yfund})]
& E_6 \;[4 \; \Yfund] & E_6 \;[3\Yfund + \overline{\Yfund}]        \nonumber\\
\downarrow  \searrow & \downarrow & \swarrow \downarrow  \nonumber\\
SO(10) (1,1,4)&F_4 \; [3 \; \Yfund] &  SO(10) (2,0,4)     \\
 \uparrow  &\searrow       \downarrow      &\swarrow  \nonumber\\
SO(11) (1,5)       &SO(9) (2,3)          & \nonumber \\
  \uparrow        &\downarrow           &\nonumber\\
SO(12) (1,0,6)          &SO(8) (2,2,2)        &\nonumber\\
\downarrow           &\downarrow           &\nonumber\\
SU(6) [6((\Yfund + \overline{\Yfund})]          &SO(7) (4,1)  &\nonumber\\
         &\downarrow  & \searrow          \nonumber\\
          &SU(4) \;[4 (\Yfund + \overline{\Yfund})]&G_2 [4 \Yfund]\nonumber
\end{eqnarray} 
\begin{eqnarray}
\label{flow5}
SO(10) (1,0,6) \longrightarrow & SO(9) (1,5) \longrightarrow &
 SO(8) (1,1,4) \nonumber\\
            &            & \swarrow  \downarrow \\
            &SO(7) (5,0) & SO(7) (2,3)          \nonumber\\
            & \downarrow &\swarrow \downarrow \nonumber\\
            &G_2 [4 \Yfund]& SU(4)\;[ 2 \Yasymm + 2(\Yfund +
 \overline{\Yfund})] \nonumber
\end{eqnarray}  
\begin{eqnarray}
\label{flow6}
SO(10) (3,0,2) & SO(10) (2,1,2)&  \nonumber\\
\downarrow     & \swarrow      &  \nonumber\\
 SO(9) (3,1)   &               &   \\
\downarrow    &                &  \nonumber\\
SO(8) (3,3,0) & SO(8)(3,2,1)   &  \nonumber\\
\downarrow  \swarrow &  \downarrow & \searrow   \nonumber\\
SO(7) (3,2) & SO(7) (4,1)      &SO(7) (5,0)\nonumber
\end{eqnarray}

\begin{eqnarray}
\label{flow7}
SO(8) (6,0,0) \longrightarrow SO(7) (0,5) \longrightarrow
 SU(4)\;[ 4 \Yasymm] \longrightarrow \mbox{Coulomb branch}
\end{eqnarray}
\begin{eqnarray}
\label{flow8}
         & SO(8) (5,1,0)& \nonumber\\
  \swarrow  &    & \searrow \nonumber\\
SO(7) (5,0) &     & SO(7) (1,4)   \\
\downarrow    & \swarrow  & \downarrow \nonumber\\
G_2 \; [4 \Yfund]&  & SU(4)\;[ 3 \Yasymm + 1(\Yfund +
 \overline{\Yfund})] \nonumber
\end{eqnarray}
\begin{eqnarray}
\label{flow9}
         & SO(8) (4,2,0)& \nonumber\\
  \swarrow  &    & \searrow \nonumber\\
SO(7) (4,1) &     & SO(7) (2,3)   \\
\downarrow    &\searrow \swarrow  & \downarrow \nonumber\\
SU(4)\;[ 4(\Yfund + \overline{\Yfund})] &G_2 \; [4 \Yfund]  &
 SU(4)\;[ 2 \Yasymm + 2(\Yfund + \overline{\Yfund})] \nonumber
\end{eqnarray}

%------------------------------------------------------------------------------
\section{New Branches} \label{branch}
%------------------------------------------------------------------------------
\subsection{Branches in  c-QMM Theories}
%------------------------------------------------------------------------------

In our previous publication we observed that c-QMM theories may have a
new branch on which the Lagrange multiplier becomes massless.  This
happens when some of the composites have infinite expectation
values. Such a scenario raises some obvious questions.  How can there
be a new massless field if some of the composites have infinite
expectation value? Infinite expectation value usually means that the
theory is weakly coupled. But at weak coupling one does not expect to
get new massless fields. The resolution to this puzzle is that the
branch actually is not at weak coupling and therefore new massless
fields are not impossible. The situation is similar to that in the
Seiberg-Witten model where new massless fields at strong coupling are
associated with a new branch \cite{SW1,SW2}.  Another issue is
the physical identification of the Lagrange multiplier. We argue it is
a massive gauge invariant field away from the new branch at infinity.

To answer these questions we look at the simple example $SU(2) \times
SU(2)$ with matter in the $(2,2)+4(2,1)+2(1,2)$ representation, which
is an s-confining theory\cite{PT}. By adding a mass to two $(2,1)$
fields and integrating these out one reaches a c-QMM theory, namely
$SU(2) \times SU(2)$ with $(2,2)+2(2,1)+2(1,2)$ matter\cite{KI}.  In
this example we are able to identify the Lagrange multipliers with
heavy mesons of the parent s-confining theory.  Thus a meson made out of
heavy quarks becomes massless at the new branch, an interesting and
puzzling phenomenon; in this connection, see \cite{DP}.

To be specific denote by $q_{\alpha\dot\alpha}$, $l_{\alpha i}$ and
$r_{\dot\alpha J}$ the matter fields transforming respectively in the
$(2,2)$, $(2,1)$ and $(1,2)$ representations of $SU(2) \times
SU(2)$. Here  $i=1,\ldots,4$ and $J=1,2$  are flavor indices, and
$\alpha$ and $\dot\alpha$ are gauge indices. The
local gauge invariant operators (mesons) are:
 
\[
M_{iJ}=q_{\alpha\dot\alpha}l_{\beta i}r_{\dot\beta
J}\epsilon^{\alpha\beta}\epsilon^{\dot\alpha\dot\beta},
\; X=q_{\alpha\dot\alpha}q_{\beta\dot\beta}
\epsilon^{\alpha\beta}\epsilon^{\dot\alpha\dot\beta},
\; L_{ij}=l_{\alpha i}l_{\beta j}\epsilon^{\alpha\beta},
\; R=r_{\dot\alpha 1}r_{\dot\beta2}\epsilon^{\dot\alpha\dot\beta},\]
\[
Y=r_{\dot\alpha 1}r_{\dot\beta2}q_{\alpha\dot\gamma}q_{\beta\dot\eta}
\epsilon^{\alpha\beta}
\epsilon^{\dot\alpha\dot\gamma}\epsilon^{\dot\beta\dot\eta}
\;\;\;\mbox{and} \;\;\; W_{ij}=l_{\alpha i}l_{\beta j}
q_{\gamma\dot\alpha}q_{\eta\dot\beta}
\epsilon^{\alpha\gamma}\epsilon^{\beta\eta}\epsilon^{\dot\alpha\dot\beta}.\]
This theory is s-confining and if the two $SU(2)$
factors have the same coupling scale $\Lambda$ the effective
superpotential is
\begin{equation}
W=-{X{\,\rm Pf\,}L-\frac14 W\cdot L\over\Lambda^3}
-{R{\,\rm Pf\,}W-\frac12 M^2\cdot W\over\Lambda^7},
\end{equation}
where $ W\cdot L=W_{ij}L_{kl}\epsilon^{ijkl}$ and $M^2\cdot
W=M_{i1}M_{j2}W_{kl}\epsilon^{ijkl}$. One may now deform this theory
by adding a mass term making two of the $l_i$ fields heavy, obtaining
at low energies  $SU(2) \times SU(2)$ with 
$(2,2)+2(2,1)+2(1,2)$, a theory with a c-QMM. From  the superpotential
\begin{equation}
W=-{X{\,\rm Pf\,}L-\frac14 W\cdot L\over\Lambda^3}
-{R{\,\rm Pf\,}W-\frac12 M^2\cdot W\over\Lambda^7}
+mL_{34}
\end{equation}
one easily finds that at any point in the moduli space the fields
$M_{iJ}$, $L_{ij}$ and $W_{ij}$, with $i$ or $j=3$ or 4 are heavy and can be
integrated out. However, special care must be exercised handling the
fields $L_{34}$ and $W_{34}$. Leaving these in, the resulting
superpotential is
\begin{equation}
\label{Wof11theory}
W={L_{34}\over\Lambda^3}[XL_{12}+W_{12}+m\Lambda^3]
+{W_{34}\over\Lambda^7}[-\Lambda^4L_{12}+RW_{12}-M^2],
\end{equation}
where $M^2=M_{11}M_{22}-M_{12}M_{21}$.  We see that at generic points
$L_{34}$ and $W_{34}$ get massive by pairing up with linear
combination of fields, but the particular combination of fields that
remains light depends on the specific point in the moduli. This is why
it is appropriate to retain the heavy fields $L_{34}$ and $W_{34}$ in
the low energy description.

Now, the superpotential of Eq.~(\ref{Wof11theory}) is appropriate for
the $SU(2) \times SU(2)$ theory with $(2,2)+2(2,1)+2(1,2)$. In the
context of this theory the fields $L_{34}$ and $W_{34}$ are Lagrange
multipliers enforcing two quantum modified constraints. A more
familiar form of the superpotential is found by inserting the $L_{34}$
constraint into the $W_{34}$ constraint, obtaining
\begin{equation}
\label{Wof11theorysimple}
W={W_{34}\over\Lambda^7}[RXL+{\rm Pf\,}\hat M+R\Lambda^4+L\Lambda^4],
\end{equation}
where $L=L_{12}$ and  $\hat M_{iJ}=M_{iJ}$
for $i,J=1,2$. We see that the Lagrange multiplier of the standard
description is ${W_{34}/\Lambda^7}$, a meson of the underlying
s-confining theory.

Although one should retain both Lagrange multipliers as in
Eq.~(\ref{Wof11theory}) to describe the whole moduli space, the
partially integrated form of Eq.~(\ref{Wof11theorysimple}) is sufficient
to identify the new branches at infinity. One only needs to set
$X=\epsilon^{-1}$, $R= L=-\epsilon\Lambda^4$ as
$\epsilon\to0$. On this trajectory $W_{34}$ is massive and the
equations of motion are satisfied only if $W_{34}=0$. But in the limit
the equations are satisfied automatically, for any $W_{34}$. On this
branch, $W_{34}$ is a massless field with massive constituents. 

At the new branch, the gauge symmetry is broken,
$SU(2)\times SU(2)\rightarrow SU(2)$. The unbroken $SU(2)$ is strongly
coupled, even though the symmetry breaking occurs at infinity on the
moduli space. In fact, as we approach the branch on the hypersurface
$W_{34}=0$ the gauge symmetry is completely broken, but with
$SU(2)\times SU(2)\rightarrow SU(2)$ at the high scale
$X=\epsilon^{-1}$ while the diagonal $SU(2)$ is broken only at the low
scale  $R= L=-\epsilon\Lambda^4$. 

We emphasize that we cannot proof the existence of this new branch and
its physical consequences. Since at the branch the theory is strongly
coupled we have no means of computing the Kahler potential, and
therefore we cannot discount the possibility that it is singular at the
new branch.

%-----------------------------------------------------------------------------
\subsection{Branches for Theories with Multiple Constraints}
%-----------------------------------------------------------------------------

For QMM-theories derived from s-confining theories the number of
constraints between the composites may be greater than one. One of the
constraints  must be quantum modified. This quantum modified
constraint arises from integrating out the composite corresponding to
the tree level mass term added. For SUSY QCD the mass term is
$W_{mass}= m(Q_{N_{f}}\overline Q_{N_{f}})$ and the corresponding
composite is the meson type operator $M_{N_{f}N_{f}}$. Integrating
this out gives the quantum modified constraint. Generally there can be
other constraints. The number of constraints is given by the formula\cite{GN}
\begin{equation}
N_{\rm Con} = N_{\rm Ops} - ( N_{\rm Fund} - {\rm dim}\, G)
\end{equation}
where $N_{\rm Fund}$, $N_{\rm Ops}$ and $N_{\rm Con}$ are the number
of fundamental chiral superfields, the number of independent gauge
invariant operators and the number of constraints, respectively, and
${\rm dim}\,  G$ is the dimension of the gauge group.
The additional constraints can be derived by integrating out other
massive composite fields from the parent s-confining theory. Because
there is no tree level mass term introduced for these fields they
generally generate constraints without quantum modifications. The
Lagrange multipliers correspond to these massive composite fields. It
is easy to see that these theories can have new branches on which the
Lagrange multipliers are massless.

As an example we consider $SU(4)$ with $[ 2 \Yasymm + 2(\Yfund +
 \overline{\Yfund})]$ which was considered in \cite{CSS}.
The matter contents and the gauge invariant operators are summarized
 in the next table:

\begin{displaymath}
\begin{array}{|l|c|ccc|c|} 
                                        \hline
      & SU(4)   &   SU(2)   &SU(2)  &  SU(2)& d.o.f.   \\    \hline
  A   & \Yasymm & \Yfund    & 1     & 1     &2 \times 6  \\
  Q   & \Yfund  & 1    &  \Yfund    & 1      &2 \times 4  \\
  \overline{Q}   &\overline{\Yfund}& 1 & 1& \Yfund &2 \times 4  \\  \hline 
M_{0}=(Q \overline{Q})  &1 & 1 &\Yfund &\Yfund & 4  \\
M_{2}=(QAA \overline{Q}) &1 &1 &\Yfund &\Yfund & 4  \\
T = (AA) &1 &\Ysymm &1 &1 &3 \\
H = (AQQ) &1 &\Yfund &1 &1 &2 \\
\overline{H} =(A\overline{Q}\,\overline{Q})  &1 &\Yfund &1 &1 &2 \\ 
\lambda &1 &1 &1 &1 &0 \\
\mu &1 &1 &1 &1 &0 \\
\hline
\end{array}
\end{displaymath}
The number of constraints adds up to  2 and the superpotential is 
\[W = \lambda[T^2 M_{0}^2- T H \overline{H}- M_{2}^2 - \Lambda^8] +
            \mu[M_{0} M_{2}+H \overline{H}]. \]
There is a new branch with a massless $\mu$. For
\[ M_{0} \sim \epsilon^{p_{1}}\; ,\; M_{2} \sim \epsilon^{p_{2}}\;,\;
    H \sim \epsilon^{p_{3}}\;,\;\overline{H} \sim \epsilon^{p_{4}}\;\;
\mbox{and}\;\; T \sim \epsilon^{-p_{1}} ,\]
$\mu$ is arbitrary in the limit $\epsilon = 0$. Here we assume that all
the $p_{i}$ are positive and $ p_{3} + p_{4}\geq p_{1}$.

One can see that this is quite generic. For example, if the
$\mu$-constraint has no linear term and there is a field in the
$\lambda$-constraint which is not in the $\mu$-constraint the $\mu$
branch is guaranteed. This new branch is reached if the T fields have
infinite field strength. This resembles the in c-QMM theories. But for
QMM-theories with multiple constraints this does not have to be
generic. One can easily imagine how one could get a new branch at
finite field strength.  For example, a superpotential of the schematic
form
\[W = \lambda[A B + C D - \Lambda^{p}] + \mu [A C E - D^2] \]
has a new branch at $\epsilon\to0$ with
\[ C \sim \epsilon^{p_{1}}, D \sim \epsilon^{p_{2}},
 E \sim \epsilon^{p_{3}} \; \mbox{and} \; p_{i} > 0\]
if $(A B) \sim \Lambda^{p}$, with  no composite field of
infinite field strength.
%-----------------------------------------------------------------------------
\section{The Quantum Modified Theories}
% ----------------------------------------------------------------------------
\subsection{$SO(N)$ Theories}
% ----------------------------------------------------------------------------
In this section we list all the theories fulfilling the index
constraint.  For each theory we list also if the theory is c-QMM and
the number of constraints.  The consistency of these tables can be
checked in different ways. A useful condition follows from the
observation that i-QMM theories can never flow into c-QMM
theories. Another, perhaps obvious, fact is that the number of
constraints can never increase while flowing from one theory down to
another.

Most of the quantum modified theories for SO gauge groups can be 
derived from s-confining theories. This is easily accomplished by using the 
tabulated s-confining theories\cite{CSS}. The gauge invariant operators 
are the same if a flavor is left out. The superpotential can 
be generated by integrating out one flavor from the s-confining 
superpotentials.

All the other quantum modified theories, those not derived from
s-confining theories, are already in the literature. The
$SO(10)(1,0,6)$ theory was first derived by Pouliot and
Strassler\cite{pouliot2}.  The results for $SO(9)(1,5)$,
$SO(8)(1,1,4)$ and $SO(7)(2,3)$ are easily generated by considering
$SO(10)(1,0,6)$ along some flat directions\cite{cho}.  The c-QMM
theory $SO(8)(0,1,5)$ was also derived by Pouliot and
Strassler\cite{pouliot2} from which $SO(7)(1,4)$ follows along flat
directions\cite{cho}. The results are summarized in
Table~\ref{SOtable}.

\begin{table}[h]
\vspace*{-4cm}
\begin{center}
\begin{tabular}{|l|l|l|l|} \hline
$SO(N)$    &  $ (0,0,N-2)$ & Coulomb branch &   \\
$SO(N)$    &  $ adjoint$   & Coulomb branch & \\
$SO(14)$   &  $ (1,0,4)$   & i-quantum modified  from s-confining theory &2  \\
$SO(13)$   &  $ (1,3)$     & i-quantum modified  from s-confining theory &2 \\
$SO(12)$   &  $ (1,0,6)$   & i-quantum modified  from s-confining theory &1 \\
$SO(12)$   &  $ (2,0,2)$   & i-quantum modified  from s-confining theory &3 \\
$SO(12)$   &  $ (1,1,2)$   & i-quantum modified  from s-confining theory &2 \\
$SO(11)$   &  $ (1,5)$     & i-quantum modified  from s-confining theory &1 \\
$SO(11)$   &  $ (2,1)$     & i-quantum modified  from s-confining theory &2 \\
$SO(10)$   &  $ (4,0,0)$   & i-quantum modified  from s-confining theory &1 \\
$SO(10)$   &  $ (3,0,2)$   & i-quantum modified  from s-confining theory &1 \\
$SO(10)$   &  $ (2,0,4)$   & i-quantum modified  from s-confining theory &1 \\
$SO(10)$   &  $ (3,1,0)$   & i-quantum modified  from s-confining theory &2 \\
$SO(10)$   &  $ (2,1,2)$   & i-quantum modified  from s-confining theory &1 \\
$SO(10)$   &  $ (1,1,4)$   & i-quantum modified  from s-confining theory &1 \\
$SO(10)$   &  $ (2,2,0)$   & i-quantum modified  from s-confining theory &2 \\
$SO(10)$   &  $ (1,0,6)$   & i-quantum modified   & 1 \\
$SO(9)$    &  $ (3,1)$     & i-quantum modified  from s-confining theory &1 \\
$SO(9)$    &  $ (2,3)$     & i-quantum modified  from s-confining theory &1 \\
$SO(9)$    &  $ (1,5)$     & i-quantum modified  & 1 \\
$SO(8)$    &  $ (5,1,0)$   & c-quantum modified  & 1 \\
$SO(8)$    &  $ (4,2,0)$   & i-quantum modified  from s-confining theory &2 \\
$SO(8)$    &  $ (3,3,0)$   & i-quantum modified  from s-confining theory &1 \\
$SO(8)$    &  $ (4,1,1)$   & i-quantum modified  & 1 \\
$SO(8)$    &  $ (3,2,1)$   & i-quantum modified  from s-confining theory &1  \\
$SO(8)$    &  $ (2,2,2)$   & i-quantum modified  from s-confining theory &1 \\
$SO(7)$    &  $ (5,0)$     & i-quantum modified  from s-confining theory &1 \\
$SO(7)$    &  $ (4,1)$     & i-quantum modified  from s-confining theory &1 \\
$SO(7)$    &  $ (3,2)$     & i-quantum modified  from s-confining theory &1 \\
$SO(7)$    &  $ (2,3)$     & i-quantum modified  & 1 \\
$SO(7)$    &  $ (1,4)$     & c-quantum modified  & 1 \\ \hline
\end{tabular}
\end{center}
\caption{These are all $SO$ theories satisfying $\sum_j \mu_j -\mu_G =
0$.  We list the gauge group and the field content of the 
theories in the first and second column. For a $SO(2N)$ theory with
$s$ spinors, $s'$ conjugate spinors and $q$ vector representations as the
matter contents we write $(s, s', q)$. For a $SO(2N+1)$ theory with
$s$ spinors and $q$ vector representations as the matter contents we
write $(s, q)$. In the third column, we indicate whether the theory 
has a quantum
modified moduli space or a Coulomb branch. The prefix ``i'' indicates 
an invariant quantum modification and the prefix ``c'' a covariant 
quantum modification. For invariant quantum modified theories we 
indicate if the quantum modified theory can be derived from a s-confining 
theory.In the last column we indicate the number of constraints.}
\label{SOtable}
\end{table}
\clearpage

%%%%%%%%%%%%%%%%%%%%%%%%%%%%%%%%%%%%%%%%%%%%%%%%%%%%%%%%%%%%%%%%%%%%%%%%%%%%%%
\subsection{Theories with Exceptional Groups }
% ----------------------------------------------------------------------------

In the following we describe all theories based on exceptional groups
which have a quantum modified moduli. To be complete we list all such
theories, even though some of them have already been introduced in the
literature\cite{cho2}. For each theory we give a table listing, in the
first column, the matter content and all the local gauge invariant
operators. The second column lists the transformation properties under
the gauge symmetry group (non-trivial only for the matter content) and
in the remaining columns we give the transformation properties under
the global ``flavor'' symmetry groups.  Lagrange multipliers are
fields and as such we include them in our lists. 

As opposed to what is done in Ref.~\cite{GN} we do not write down the
explicit contraction of all the indices for the gauge invariant
operators. Detailed analysis like that of Ref.~\cite{GN} is necessary
in order to write down the exact form of the classical constraints,
which can be checked explicitly. This was useful for some involved
cases there, but the behavior of the theories in this paper can be
determined without resorting to such detailed analysis. While we
cannot write down the superpotentials, we give under each table the
degree of homogeneity of the superpotential with respect to each
matter field.  

To find the schematic form of  a superpotential one simply forms all the
flavor and gauge invariant operators out of the listed gauge invariant
operators. The terms appearing in the superpotential are only
constrained by the fixed power under which each fundamental field must
appear (which, as stated above, is listed under each table). We
derived these by flowing down to known theories.  We checked 't~Hooft
anomaly matching for all the theories.

Table~\ref{exctable} summarizes the findings for exceptional groups.
%------------------------------------------------------------------
\begin{table}[h]
%\vspace*{-1cm}
\begin{center}
\begin{tabular}{|l|l|l|l|} \hline
$E_8$    &  $ \Yfund = adjoint $             & Coulomb branch &   \\
$E_7$    &  $ adjoint          $             & Coulomb branch &   \\
$E_7$    &  $ 3 \Yfund         $             & i-quantum modified &1\\
$E_6$    &  $ adjoint         $              & Coulomb branch  &  \\
$E_6$    &  $ 4 \Yfund         $             & i-quantum modified &1\\
$E_6$    &  $ 3 \Yfund + \overline{\Yfund}$  & i-quantum modified &1\\
$E_6$    &  $ 2 (\Yfund + \overline{\Yfund})$ &i-quantum modified &1\\
$F_4$    &  $ adjoint         $               &Coulomb branch  &  \\
$F_4$    &  $ 3 \Yfund         $              &i-quantum modified &1\\ 
$G_2$    &  $ 4 \Yfund        $  & i-quantum modified from s-confining &1 \\
$G_2$    &  $ adjoint          $             & Coulomb branch  &  \\  \hline
\end{tabular}
\end{center}
\caption{These are all exceptional theories satisfying $\sum_j \mu_j -\mu_G =
0$.  We list the gauge group and the field content of the 
theories in the first and second column.  In the third column, we indicate 
whether the theory has a quantum modified moduli space or a  
Coulomb branch. The prefix ``i'' indicates 
an invariant quantum modification and the prefix ``c'' indicates a covariant 
quantum modification. Only one of the theories can be derived from s-confining 
theories.In the last column we indicate the number of constraints.}
\label{exctable}
\end{table}

%------------------------------------------------------------------------------

%------------------------------------------------------------------------------
\subsubsection{$F_4$ with $ 3 \protect\Yfund $}\label{F_4}

\begin{displaymath}
\begin{array}{|l|c|cc|} 
                                        \hline
       &  F_4 &   SU(3)      &  U(1)_{R}   \\    \hline
  Q    & \Yfund  &  \Yfund  &  0          \\  \hline
  Q^2  & 1       &  \Ysymm  &  0          \\
  Q^3  & 1       &  \Ythrees  &  0          \\
  Q^4  & 1       &  \Ysquare  &  0          \\
  Q^5  & 1       &  \overline{\Yfund}  &  0          \\
  Q^6  & 1       &  1  &  0          \\
  Q^9  & 1       &  1  &  0          \\
 \lambda& 1       & 1  &  2          \\
\hline
\end{array}
\end{displaymath}

$Q$ appears as $Q^{18}$ in the superpotential.
\clearpage
%%%%%%%%%%%%%%%%%%%%%%%%%%%%%%%%%%%%%%%%%%%%%%%%%%%%%%%%%%%%%%%%%%%%%%%%%%%%%%

\subsubsection{$E_6$ with $ 2( \protect\Yfund + \overline{\protect\Yfund})$}\label{E_6a}

\begin{displaymath}
\begin{array}{|l|c|ccc|} 
                                        \hline
            &  E_6 &   SU(2) & SU(2)     &  U(1)_{R}   \\    \hline
  Q         & \Yfund  &  \Yfund  &  1  &0                  \\  
\overline{Q}& \overline{\Yfund} & 1 & \Yfund  &  0     \\    \hline
  Q\overline{Q} & 1 &\Yfund &\Yfund  &  0          \\
  Q^{3} & 1 &\Ythrees & 1  &  0          \\
  \overline{Q}^{3} & 1 & 1  &\Ythrees  &  0          \\
  Q^{2}\overline{Q}^{2} & 1 &\Ysymm & \Ysymm &  0          \\
  Q\overline{Q}^{4} & 1 &\Yfund &1  &  0          \\
  Q^{4}\overline{Q} & 1 & 1  &\Yfund  &  0          \\
  Q^{3}\overline{Q}^{3} & 1 &\Yfund &\Yfund  &  0          \\
  Q^{4}\overline{Q}^{4} & 1 &1 & 1 &  0          \\
  Q^{6}\overline{Q}^{6} & 1 &1 & 1 &  0          \\
\lambda                 & 1 &1 & 1 &  2          \\
\hline
\end{array}
\end{displaymath}

$Q$ appears as $Q^{12}$ in the superpotential and $\overline{Q}$ as
 $\overline{Q}^{12}$.
%%%%%%%%%%%%%%%%%%%%%%%%%%%%%%%%%%%%%%%%%%%%%%%%%%%%%%%%%%%%%%%%%%%%%%%%%%%%%%

\subsubsection{$E_6$ with $ 3 \protect\Yfund + \overline{\protect\Yfund} $}\label{E_6b}

\begin{displaymath}
\begin{array}{|l|c|ccc|} 
                                        \hline
       &  E_6     &   SU(3)      &  U(1)_{A} &  U(1)_{R}   \\    \hline
  Q    & \Yfund  &  \Yfund      &  1        & 0           \\ 
\overline{Q}& \overline{\Yfund} & 1 &-3     &  0          \\    \hline
 Q\overline{Q} & 1 & \Yfund & -2 &  0          \\
 Q^{3} & 1 &\Ythrees & 3  &  0          \\
 \overline{Q}^{3} & 1 &1 &-9  &  0          \\
 Q^{2}\overline{Q}^{2} & 1 &\Ysymm &-4  &  0          \\
 Q^{4}\overline{Q} & 1 & \Ysquare& 1 &  0          \\
 Q^{6}   & 1 & 1 &  1  &  0          \\
 Q^{5}\overline{Q}^{2} & 1 & \Yfiveboxes & -1 &  0          \\
 Q^{9}\overline{Q}^{3} & 1 & 1& 0 &  0          \\
 \lambda& 1       & 1  &  0 &  2          \\
\hline
\end{array}
\end{displaymath}

$Q$ appears as $Q^{18}$ in the superpotential and $\overline{Q}$ as
 $\overline{Q}^{6}$.
%%%%%%%%%%%%%%%%%%%%%%%%%%%%%%%%%%%%%%%%%%%%%%%%%%%%%%%%%%%%%%%%%%%%%%%%%%%%%%

\subsubsection{$E_6$ with $ 4 \protect\Yfund $}\label{E_6c}

\begin{displaymath}
\begin{array}{|l|c|cc|} 
                                        \hline
       &  E_6 &   SU(4)      &  U(1)_{R}   \\    \hline
  Q    & \Yfund  &  \Yfund  &  0          \\  \hline
  Q^3  & 1       &  \Ythrees  &  0          \\
  Q^6  & 1       &  \Ysymtwoathree  &  0          \\
  Q^{12} & 1       &  1  &  0          \\
\lambda& 1       & 1  &  2          \\
\hline
\end{array}
\end{displaymath}

$Q$ appears as $Q^{24}$ in the superpotential.
%%%%%%%%%%%%%%%%%%%%%%%%%%%%%%%%%%%%%%%%%%%%%%%%%%%%%%%%%%%%%%%%%%%%%%%%%%%%%%

\subsubsection{$E_7$ with $ 3 \protect\Yfund $}\label{E_7}

\begin{displaymath}
\begin{array}{|l|c|cc|} 
                                        \hline
       &  E_7 &   SU(3)      &  U(1)_{R}   \\    \hline
  Q    & \Yfund  &  \Yfund  &  0          \\  \hline
  Q^2  & 1       &  \Yasymm  &  0          \\
  Q^4  & 1       &  \Yfours  &  0          \\
  Q^6  & 1       &  \Ysymthreeatwo  &  0          \\
  Q^8  & 1       &  \Ysymm  &  0          \\
  Q^{12}  & 1       &  1  &  0          \\
  Q^{18}  & 1       &  1  &  0          \\
 \lambda& 1       & 1  &  2          \\
\hline
\end{array}
\end{displaymath}

$Q$ appears as $Q^{36}$ in the superpotential.
%%%%%%%%%%%%%%%%%%%%%%%%%%%%%%%%%%%%%%%%%%%%%%%%%%%%%%%%%%%%%%%%%%%%%%%%%%%%%%
\section{Conclusions}
Our results are summarized in Tables~\ref{SOtable} and
\ref{exctable}. They present all SUSY field theories based on a simple
orthogonal or exceptional gauge group that satisfy the index
constraint $\sum\mu_i-\mu_G=0$. These theories may either have a
quantum modified moduli  (QMM) space or a Coulomb branch with massless
photons, and the right alternative is listed in the tables. For
theories with a QMM we give the finer distinction of whether the constraint
that is quantum modified is invariant (i-QMM) or only covariant (c-QMM)
under the global symmetries of the theory. Finally, the tables list
the number of constraints needed to specify the QMM.

This work completes the classification begun in Ref.~\cite{GN}, which dealt
with unitary and symplectic simple gauge groups. In that work it was
noted that every i-QMM theory flowed from an s-confining theory and it
was shown that every s-confining theory flows to a theory with an
i-QMM or a Coulomb branch. However, we have seen in this paper that
there exist theories (exactly four) with an i-QMM which however cannot
be obtained by flowing from s-confining theories.

Our work complements recent efforts\cite{CS2,GN,DMS} to understand the
behavior of all SUSY gauge theories based on a simple gauge group with
matter satisfying $\sum\mu_i-\mu_G\le0$.

As in Ref.~\cite{GN} we note that theories with a c-QMM exhibit a peculiar
new branch at a point on the boundary of the moduli space. Here we
have elucidated some of the physics of this branch by considering
this theory as a deformation of an s-confining theory. The Lagrange
multiplier that becomes massless on the branch corresponds to a
composite of heavy constituents! This is possible because the theory
remains strongly coupled on this branch.

\vskip1.2cm
{\it Acknowledgments}
\hfil\break
We are grateful to Ken Intriligator, Erich Poppitz and Witold Skiba
for many helpful discussions, particularly with regard to the
$SU(2)\times SU(2)$ theories.  This work is supported by the
Department of Energy under contract DOE-FG03-97ER40506.
%%%%%%%%%%%%%%%%%%%%%%%%%%%%%%%%%%%%%%%%%%%%%%%%%%%%%%%%%%%%%%%%%%%%%%%%%
\clearpage
\appendix

%%%%%%%%%%%%%%%%%%%%%%%%%%%%%%%%%%%%%%%%%%%%%%%%%%%%%%%%%%%%%%%%%%%%%%%%%%%%%%
\section{Appendix}
\subsection{Flow of the Exceptional Groups}
In order to determine whether a theory has a QMM one must explore the
whole moduli space. It is possible that a theory satisfying the index
constraint $\sum_{i=1}^{n} \mu_{i} - \mu_{G} = 0$ has a Coulomb branch
or a branch with $\sum_{i=1}^{n} \mu_{i} - \mu_{G} \neq 0$. The whole
moduli space must be explored because the pattern of symmetry breaking
generally dictates into which  different branches the theory flows.

Finding the full moduli space of the exceptional groups, however, is
difficult. Given the matter content it is useful to have at hand all
possible patterns of symmetry breaking. In this appendix we describe a
simple, general method for determining all possible patterns of
symmetry breaking given a gauge group and matter content.

It will be useful to have an illustrative example at hand. Consider an
$SU(3)$ theory with matter $A$ in the adjoint representation. There
are two patterns of symmetry breaking possible, given by the
expectation values
\[  \langle A\rangle= \left( \begin{array}{ccc}
                    1 & 0 & 0 \\
                    0 & 1 & 0 \\
                    0 & 0 & -2 \\
 \end{array} \right)  \]
or 
\[ \langle A\rangle= \left( \begin{array}{ccc}
                    1 & 0 & 0 \\
                    0 & -1 & 0 \\
                    0 & 0 & 0 \\
 \end{array} \right). \]
$SU(3)$ is broken to $SU(2) \times U(1)$ or $U(1) \times U(1)$
respectively.  Note that for $SU(N)$ with fundamental matter only the
pattern of symmetry breaking is fixed because an $SU(N)$
transformation can bring fundamentals to a standard form.  However,
for exceptional groups a unique standard form for the fundamentals
does not exist. Even if we had a concrete representation of the
exceptional groups (for example $26 \times 26$ matrices for $F_4$) it
would still be difficult to determine all the breaking patterns.
Therefore we exploit a method based on roots and weights. The
advantage is that starting from a Dynkin diagram one can determine
the roots and weights in a finite number of steps\cite{hum,geo}, and then our
analysis uses an additional finite number of steps. Therefore our
method is programmable.

Therefore, to analyze the patterns of symmetry breaking we abandon
tensor notation for the Higgs fields and return to the more basic
representation in terms of weights. Recall that generators
($E_\alpha$) are labeled by roots ($\alpha$) and act on weights
($\mu$) like raising and lowering operators ($E_\alpha$ and
$E_\alpha^\dagger=E_{-\alpha}$).  The idea is to give a vacuum
expectation value (VEV) directly to a weight $\mu_{i}$ or to a linear
combination of some weights. To see how the generators act on the
Higgs corresponding to the weight $\mu_{i}$ we just determine how the
roots act on the weight $\mu_{i}$. By construction the action of
roots on weights is restricted to motions within the weight lattice of
the corresponding representation. Either the root moves the weight to
some other weight or annihilates it. 

It would seem that the solution to our problem is at hand, because
given a weight with a VEV the roots that do not annihilate it
correspond to broken generators. A slightly modified version of this
statement and its converse --- that unbroken generators correspond to
roots that annihilate the weight --- are true. There are  two subtleties:
\begin{itemize}

\item The operators $E_{\pm\alpha}$ are not hermitian. The pattern of
symmetry breaking is determined from the hermitian generators
constructed from them,
\( T_{+} = E_{\alpha} + E_{- \alpha}\) and \( T_{-} = i(E_{\alpha} 
- E_{- \alpha})\). If at least one of the two operators
$E_{\pm\alpha}$ is broken then both $T_{+}$ and $T_{-}$ are broken.

\item If two roots move the weight $\mu_i$ with a VEV to the same
weight $\mu_j$ then a linear combination of roots remains
unbroken. This occurs when the roots are at the origin, ie, when the
generators are in the Cartan subalgebra. It also may happen if the VEV
is given to a linear combination of weights.
\end{itemize}

In sum, given a VEV to a weight $\mu_i$ and a root $\alpha$ that moves
the weight to $\mu_j\neq\mu_i$, then both $E_\alpha$ and $E_{-\alpha}$
are broken generators. If there are $n$ elements of the Cartan
subalgebra then one can always find $n-1$ unbroken linear
combinations. If a VEV is given to a linear combination of weights and
$n$ roots map this combination to the same weight, then only $n-1$ of
the corresponding generators are unbroken.\footnote{
Determining these combinations can prove difficult. Acting with roots on
weights is only defined up to  phases.  The only requirement on these
phases is that the whole algebra be consistent. If, for example, two
products of simple roots move from $\mu_{i}$
to $\mu_{j}$
\[ E_{\alpha_1}E_{\alpha_2}\cdots E_{\alpha_n} \mu_{i} = 
e^{i \; \phi_{1}} \mu_{j} 
\;\;\; \mbox{and}\;\;\;
 E_{\beta_1}E_{\beta_2}\cdots E_{\beta_n} \mu_{i} = 
e^{i \; \phi_{2}} \mu_{j} .\] 
the phases $\phi_1$ and $\phi_2$ are not necessarily equal. 
There are general ways how to determine the phases. But for the
following examples we do so  on a case by case basis.}

While the considerations above apply to any field theory, for SUSY
theories there are further simplifications. If the gauge symmetry of a
SUSY theory is $G$, then the potential is in fact invariant under a
larger symmetry, namely the complexified group $G_c$. If the symmetry
is spontaneously broken $G\to H$, then the potential of the low energy
theory has $H_c$ symmetry. Therefore, in a supersymmetric vacuum both
lowering and raising operators must be broken
simultaneously.\footnote{However, the reverse is not true: even if
both raising and lowering operators are broken simultaneously the
vacuum may not be supersymmetric. One must in addition verify that the
$D$-flatness conditions are satisfied.}

Let us return to our illustrative example, $SU(3)$ with adjoint
matter.  The root diagram is
\[ \begin{array}{lrclr}
  (-1/2, \sqrt{3}/2)& & & & (1/2,\sqrt{3}/2)  \\
      & \nwarrow & & \nearrow &              \\
 (-1,0)&  \longleftarrow & (0,0)_{1},(0,0)_{2} & \longrightarrow &  (1,0)  \\
      & \swarrow  &    & \searrow   &        \\
 (-1/2, -\sqrt{3}/2)& & & & (1/2, -\sqrt{3}/2) .
\end{array}  \]
In this example the weight diagram of the matter field is  exactly the
same.  

Given, for example, a VEV to  $|(1/2,\sqrt{3}/2)\rangle$ 
the roots that annihilate it are
\[((0,0)_{1}-1/\sqrt{3}(0,0)_{2} ),\;(1,0) , \;(1/2,\sqrt{3}/2) , \;
 (-1/2, \sqrt{3}/2) ,\]
but only the first gives an unbroken generator, because the last three
have hermitian conjugates which are broken. The symmetry is
broken down to $U(1)$. The alert reader may be concerned about adjoint
breaking to a non-maximal subgroup. In this example we have given a
VEV to a complex weight, which cannot be accomplished with a single
hermitian matrix of scalars  $A$. However, two scalar fields $A$ and
$B$ will accomplish this, and it is clear that unaligned VEVs to two
adjoints can break $SU(3)$ to $U(1)$. It is also clear that this vacuum
is not supersymmetric. This follows from the general considerations
above, but one can easily check that the D-flatness condition is not
satisfied. 

If we give a VEV to the real combination $|(1/2,\sqrt{3}/2)\rangle+
|(-1/2,-\sqrt{3}/2)\rangle$ the unbroken operators are
\[(0,0)_{1}-1/\sqrt{3}(0,0)_{2} , \;(1/2,\sqrt{3}/2)+(-1/2,-\sqrt{3}/2) .\]
The pattern of symmetry breaking is $SU(3)\rightarrow U(1) \times
U(1)$. By symmetry, giving a VEV to any other non zero root results in
the same breaking pattern.

Consider a VEV of the zero roots. For  $|(0,0)_{1}\rangle $  the
unbroken operators are
\[(0,0)_{1}, (0,0)_{2} \]
and for  $|(0,0)_{2}\rangle $ they are
\[(0,0)_{1},(-1,0),(1,0), (0,0)_{2} .\]
The first set corresponds to $U(1) \times U(1)$ the second to $SU(2)
\times U(1)$.  

In the following sections we give further examples illustrating the
method with exceptional groups. Although not presented here, we have
used this method to check the symmetry breaking patterns of SO-groups
with spinors.
%%%%%%%%%%%%%%%%%%%%%%%%%%%%%%%%%%%%%%%%%%%%%%%%%%%%%%%%%%%%%%%%%%%%%%%%%%%%%% 
\subsubsection{$F_4$ with $ \protect\Yfund $}

The 52 roots of $F_4$ are
\[ \begin{array}{lrcl}
 \pm e_{i} \pm e_{j} & (24) & \mbox{with} & i\neq j \; \mbox{and} \;
i,j=1,2,3,4 \\
\pm e_{i} & (8)  & \mbox{with} & i=1,2,3,4  \\
1/2(\pm e_{1} \pm e_{2} \pm e_{3} \pm e_{4}) & (16) &\\
(0,0,0,0)_{i} & (4) & \mbox{with} & i=1,2,3,4.
\end{array}  \]
and the 26 weights of the fundamental representation are
\[ \begin{array}{lrcl}
 \pm e_{i} \pm e_{j} & (24) & \mbox{with} & i\neq j \; \mbox{and} \;
i,j=1,2,3,4 \\
(0,0,0,0)_{a} & (2) & \mbox{with} & a=1,2.
\end{array}  \]
Here $(e_i)_j=\delta_{ij}$ is the standard 4-dimensional basis of
orthonormal vectors, and the numbers in parenthesis denote the number
of such roots and weights.  Giving a VEV to $(|(e_{1} + e_{2})\rangle
+ |-(e_{1} + e_{2})\rangle )$ breaks the gauge group down to
$SO(8)$. But if we give the VEV to the zeros we obtain $SO(8)$ or
$SO(9)$. 

%%%%%%%%%%%%%%%%%%%%%%%%%%%%%%%%%%%%%%%%%%%%%%%%%%%%%%%%%%%%%%%%%%%%%%%%%%%%%%%
\subsubsection{$E_6$ with $ \protect\Yfund + \overline{\protect\Yfund} $}

The 78 roots of $E_6$ are
\[ \begin{array}{lrcl}
 \pm e_{i} \pm e_{j} & (40) & \mbox{with} & i\neq j \; \mbox{and} \;
i,j=1,2,3,4,5 \\
1/2(\pm e_{1} \pm e_{2} \pm e_{3} \pm e_{4} \pm e_{5} \pm \sqrt{3} e_{6}) &
 (32) & \mbox{with} & \mbox{even number of minus signs}\\
(0,0,0,0)_{i} & (6) & \mbox{with} & i=1,2,3,4,5,6.
\end{array}  \]
The 27 weights of the fundamental representation are
\[ \begin{array}{lrcl}
2/\sqrt{3} e_{6} & (1) &  & \\
 \pm e_{i} - 1/\sqrt{3} e_{6} & (10) & \mbox{with} & i=1,2,3,4,5 \\
1/2(\pm e_{1} \pm e_{2} \pm e_{3} \pm e_{4} \pm e_{5} + 1/\sqrt{3} e_{6}) &
 (16) & \mbox{with} & \mbox{odd number of minus signs}.
\end{array}  \]
The weights of the antifundamental representation are the negative
of the weights of the fundamental representation. 

Now give a VEV to $|(2/\sqrt{3} e_{6})\rangle $. The roots that
annihilate it are
\[ \pm e_{i} \pm e_{j}, \; \mbox{5 zero roots and } \; 
1/2(\pm e_{1} \pm e_{2} \pm e_{3} \pm e_{4} \pm e_{5} - \sqrt{3} e_{6}) .\]
The last 16 do not give unbroken generators. The unbroken roots
precisely correspond to the roots of $SO(10)$. In a SUSY theory this
pattern of symmetry breaking requires in addition an antifundamental
VEV to  $|(-2/\sqrt{3} e_{6})\rangle $
(for D-flatness).

There are alternative breaking patterns. If one gives a  VEV to 
$|(2/\sqrt{3} e_{6})\rangle  + |(e_{5} - 1/\sqrt{3} e_{6})\rangle  +
|(-e_{5} - 1/\sqrt{3} e_{6})\rangle $  then $E_6$
breaks to $F_4$.

%%%%%%%%%%%%%%%%%%%%%%%%%%%%%%%%%%%%%%%%%%%%%%%%%%%%%%%%%%%%%%%%%%%%%%%%%%%%%%
\subsubsection{$E_7$ with $ \protect\Yfund $}

The 133 roots are
\[ \begin{array}{lrcl}
 \pm e_{i} \pm e_{j} & (60) & \mbox{with} & i\neq j \; \mbox{and} \;
i,j=1,2,3,4,5,6 \\
1/2(\pm e_{1} \pm e_{2} \pm e_{3} \pm e_{4} \pm e_{5} \pm e_{6} 
\pm \sqrt{2} e_{7}) & (64) & \mbox{with} & \mbox{odd number of minus signs for
$e_{1}$ to $e_{6}$}\\
\pm \sqrt{2} e_{7} & (2) &  &  \\
(0,0,0,0)_{i} & (7) & \mbox{with} & i=1,2,3,4,5,6,7
\end{array}  \]
and the 56 weights of the fundamental representation are
\[ \begin{array}{lrcl}
 \pm e_{i} \pm 1/\sqrt{2} e_{7} & (24) & \mbox{with} & i=1,2,3,4,5,6 \\
1/2(\pm e_{1} \pm e_{2} \pm e_{3} \pm e_{4} \pm e_{5} \pm e_{6} ) &
 (32) & \mbox{with} & \mbox{even  number of minus signs}. 
\end{array}  \]
Giving a VEV to $|1/2( e_{1} + e_{2} + e_{3} + e_{4} +
 e_{5} + e_{6} )\rangle  + |- 1/2( e_{1} + e_{2} + e_{3} + e_{4} +
 e_{5} + e_{6} )\rangle $  breaks $E_7$ to $E_6$.

%%%%%%%%%%%%%%%%%%%%%%%%%%%%%%%%%%%%%%%%%%%%%%%%%%%%%%%%%%%%%%%%%%%%%%%%%%%%%%

%%%%%%%%%%%%%%%%%%%%%%%%%%%%%%%%%%%%%%%%%%%%%%%%%%%%%%%%%%%%%%%%%%%%%%%%%%%%%%

\end{document}